\documentclass[reprint,groupedaddress,superscriptaddress,amsmath,amssymb,aps]{revtex4-2}
\usepackage{graphicx}
\usepackage{dcolumn}
\usepackage{bm}
\usepackage{bbold}
\usepackage{braket}

\usepackage[usenames,dvipsnames]{xcolor}
\usepackage{textcomp}
\usepackage[toc, page]{appendix}
\usepackage{eucal}
\usepackage{caption}
\usepackage{subcaption}
\usepackage{epstopdf}
\usepackage{float}
\usepackage{svg}
\usepackage{xpatch}
\usepackage{eqnarray, amsmath}
\usepackage{amsthm}
\usepackage{scalerel}
\usepackage{stackengine,wasysym}
\usepackage{yfonts}
\usepackage{bbm}
\usepackage{mathbbol}
\usepackage{amsfonts}
\usepackage{dsfont}
\usepackage{xr-hyper}
\usepackage[colorlinks,citecolor=Blue,linkcolor=Red, urlcolor=Blue]{hyperref}
\makeatletter
\newcommand*{\addFileDependency}[1]{
\typeout{(#1)}
\@addtofilelist{#1}
\IfFileExists{#1}{}{\typeout{No file #1.}}
}\makeatother

\newcommand*{\myexternaldocument}[1]{%
\externaldocument{#1}%
\addFileDependency{#1.tex}%
\addFileDependency{#1.aux}%
}

\myexternaldocument{Supp_Mat_LF18802}

\newcommand{\be}{\begin{equation}}
\newcommand{\ee}{\end{equation}}
\newcommand{\bqa}{\begin{eqnarray}}
\newcommand{\eqa}{\end{eqnarray}}

\newcommand{\flo}[1]{{\color[rgb]{0.2,0.3,0.6}{#1}}}

\begin{document}

\title{Quantum simulations of time-dependent Hamiltonians beyond the quasi-static approximation}
\author{Boyuan Shi}
\affiliation{Blackett Laboratory, Imperial College London,
London SW7 2AZ, United Kingdom}
\author{Florian Mintert}
\affiliation{Blackett Laboratory, Imperial College London,
London SW7 2AZ, United Kingdom}
\affiliation{Helmholtz-Zentrum Dresden-Rossendorf, Bautzner Landstraße 400, 01328 Dresden, Germany}

\begin{abstract}
Existing approaches to analogue quantum simulations of time-dependent quantum systems rely on perturbative corrections to quantum simulations of time-independent quantum systems.
We overcome this restriction to perturbative treatments with an approach based on flow equations and a multi-mode Fourier expansion. The potential of the quantum simulations that can be achieved with our approach is demonstrated with the pedagogical example of a Lambda-system and the quench in finite time through a quantum phase transition of a Chern insulator in a driven non-interacting Hubbard system.
The example of the Lambda-system demonstrates the ability of our approach to describe situations beyond the validity of adiabatic approximations.
\end{abstract}
\maketitle
\section{Introduction}
There is an abundance of open questions in quantum physics, that we will most likely not be able to solve with classical computational means. Only the use of quantum simulators seems to allow us to overcome the computational complexity of many quantum mechanical many-body problems~\cite{quantum_simulation_roadmap,Quantum_Simulation_Review}.

The hardware that is required to accurately mimic the dynamics induced by a given Hamiltonian is sufficiently advanced to use quantum simulators for problems that are outside the reach of classical computational hardware.
Notable platforms include atomic gases in optical lattices \cite{quantum_simulation_ultra_cold}, crystals of trapped ions \cite{ion_crystal_2016}, arrays of Rydberg atoms~\cite{Antoine_Browaeys, 256_Lukin, 219_Lukin}, and superconducting qubits \cite{super_conducting_qubits}
that can be used to quantum simulate strongly interacting Hubbard models~\cite{quantum_simulation_ultra_cold}, topologically non-trivial phases of matter~\cite{Nature_Esslinger_Haldane}, interacting quantum spin models~\cite{quantum_simulation_spin_models, ion_crystal_2016} and quantum chemistry~\cite{quantum_simulation_quantum_Chemistry}.
While many such problems are defined in terms of a time-independent Hamiltonian, there is also a broad range of problems resultant from time-dependent Hamiltonians,
such as laser-driven dynamics of electrons in molecules~\cite{atto_physics},
quenches across boundaries between quantum phases~\cite{Kun_Yang}, time crystals~\cite{discrete_TC}, diabatic switching between different Hamiltonians~\cite{FT_Peter, FT_Jarzynski} or cycles of quantum thermodynamical machines~\cite{quantum_thermal}.

The theory and experiments on quantum simulation, so far, have mostly focused on time-independent Hamiltonians~\cite{floquet_andre,floquet_eingeering_takashi}.
A crucial reason for this restriction in theoretical work is the rigorous footing that Floquet theory provides for the definition of an effective, time-independent Hamiltonians, whereas the definition of an effective time-dependent Hamiltonian in a driven quantum system is much more problematic.

While generalizations of the Floquet theorem to aperiodically driven systems have proven difficult to find,
they are also not necessary for purposes of time-dependent quantum simulations.
An effective Hamiltonian can be defined in terms of any finite interval of system dynamics, and flow equations~\cite{Wegner, AAF_PRL_2013, Vogl_PRX} provide a solid basis for this~\cite{PRA_Vilnius_2022}.
In particular, they allow to ensure that the system dynamics is covered exactly by the effective Hamiltonian at periodic instances, even though the time-dependence of the actual system has no such periodicity.

Despite the solid foundations that the flow equations provide for the definition of a time-dependent effective Hamiltonian~\cite{Malz_and_Smith, PRA_Vilnius_2022},
any practical construction requires a separation of time-scales, with fast time-dependencies resulting in desired effective processes, and a slow time-scale for the time-dependence of the effective Hamiltonian.
This, in turn, implies either very fast and strong driving or long duration of an experiment.
The former unavoidably induces undesired processes, such as heating in the case of atomic gases~\cite{heating_1,heating_2,heating_3}, or leakage beyond the levels that define individual qubits~\cite{leakage_1,leakage_3},
and the latter typically results in conflicts with coherence time~\cite{decoherence}.

The goal of this paper is thus to develop a framework that allows for quantum simulations of time-dependent quantum systems without the requirement of such a separation of time scales.

\section{\flo{The formal framework}}
Hardly any Hamiltonian that one would want to quantum simulate can be exactly realized experimentally. It is rather necessary to realize a time-dependent Hamiltonian $H(t)$ that induces dynamics which approximates the dynamics of interest. The central, underlying mechanism is the fact that a Hamiltonian
\begin{equation}
H_{U}(t)=U(t)H(t)U^{\dagger}(t)-iU(t)\dot U^\dagger(t)
\label{eq:deco}
\end{equation}
defined in terms of a time-dependent unitary transformation $U(t)$ can describe different physics than the underlying Hamiltonian $H(t)$.
It can thus be possible to realize a Hamiltonian in the frame defined by $U(t)$, even if this Hamiltonian is practically out of reach in the laboratory frame~\cite{Eckhardt_2005,Bukov_Luca_Anatoli, Nature_Esslinger_Haldane}.

The propagator $V_U(t)$ induced by $H_U(t)$ reads $V_U(t)=U(t)V_H(t)$ in terms of the propagator $V_H(t)$ induced by $H(t)$.
Since the propagators $V_U(t)$ and $V_H(t)$ differ by a factor $U(t)$, it is essential for $U(t)$ to reduce to the identity ${\mathbbm 1}$ when observations are being taken.
The identification of a transformation $U(t)$ that achieves this reduction periodically is formalized in terms of the Floquet theory for time-independent effective Hamiltonians~\cite{floquet_andre, Dalibard_PRX}.
In the case of aperiodically driven systems, the unitary $U(t)$ will typically not be periodic, but it is essential that it reduces to the identity at well-defined points in time.
This can be achieved with the framework of flow equations~\cite{Vogl_PRX, PRA_Vilnius_2022} that considers a family of unitaries $U_s(t)$ parametrised by a parameter $s$ in terms of the differential equation $\frac{\partial U_s(t)}{\partial s}=i\eta_s(t)U_s(t)$ with an hermitian generator $\eta_s(t)$.
Associated with each such unitary $U_s(t)$ is a Hamiltonian $H_s(t)$ (following Eq.~\eqref{eq:deco})
that satisfies the differential equation~\cite{AAF_PRL_2013,Vogl_PRX}
\begin{equation}
    \frac{\partial H_s(t)}{\partial s}=i[\eta_s(t), H_s(t)]-\frac{\partial\eta_s(t)}{\partial t}\ .
    \label{eq:flow}
\end{equation}
The generator $\eta_s(t)$ is typically an explicit function of the flowing Hamiltonian $H_s(t)$, such that Eq.~\eqref{eq:flow} is actually non-linear in $H_s(t)$.
It typically has a stationary solution in the limit $s\to\infty$, and the flowing Hamiltonian $H_s(t)$ with the initial condition $H_{s=0}=H(t)$ approaches the effective Hamiltonian in this limit.

Through the explicit choice of generator $\eta_s(t)$ one can specify general properties that the effective Hamiltonian shall have. While, typically the effective Hamiltonian is expected to be time-independent, we will require that the effective Hamiltonian does not have any time-dependence associated with a fundamental driving frequency, but that it can still have time-dependence associated with some other frequencies.

The definition of the generator $\eta_s(t)$ is facilitated in term of a multi-mode Fourier series~\cite{multi_mode_Fourier, Martin_Refael_Halperin, Multi_Mode_Prethermalization, Verdeny_Puig_Mintert}  
$H_s(t)=\sum_{\bm{m}} h_s^{\bm{m}}\ e^{i\bm{m}\cdot\bm{\omega}t}$ for the flowing Hamiltonian $H_s(t)$,
where $\bm{m}$ is a vector of integers, $\bm{\omega}$ is a vector of mutually incommensurate frequencies ({\it i.e.} frequencies whose ratios are not rational bumbers), and the operators $h_s^{\bm{m}}$ are generalized Fourier coefficients.
The number of frequencies, {\it i.e.} the dimension of $\bm{m}$ and $\bm{\omega}$ can be chosen in accordance with the problem to-be-quantum-simulated; subsequent examples in this paper are based on driving with two fundamental frequencies.

The generator
\begin{equation}
\eta(s,t)=-\frac{i}{\omega_{1}}\sum_{\bm{m}} f_{\bm{m}}\ h_s^{\bm{m}}\ e^{i\bm{m}_0\cdot\bm{\omega}_{0}t}\ (e^{im_{1}\omega_{1}t}-1)\ ,
\label{eq:generator}
\end{equation}
with $\bm{m}_0=[0,m_2,m_3,\hdots]$, $\bm{\omega}_{0}=[0,\omega_{2},\omega_{3},\hdots]$
and real scalars $f_{\bm{m}}$ contains the operators $h_s^{\bm{m}}$ and is thus a function of the flowing Hamiltonian $H_s(t)$.
The factor $(e^{im_{1}\omega_{1}t}-1)$ ensures that the generator $\eta(s,t)$ vanishes at any time $t$ that is an integer multiple of the period $2\pi/\omega_1$.
This, in turn, guarantees that the unitary $U(t)$ in Eq.~\eqref{eq:deco} periodically reduces to the identity.

Besides the condition $f_{\bm{m}}=-f_{-\bm{m}}$ that ensures that the generator $\eta_s(t)$ is hermitian,
and the condition $f_{{\bm m}_0}=0$ required for convergence, there is substantial freedom in the choice of the scalars $f_{\bm{m}}$, and this freedom of choice can be used to specify which of the frequencies in $\bm{\omega}$ are contained in the effective Hamiltonian, and which frequencies are meant to play the role of driving in order to realize effective processes.
In the following, we will use $\omega_1$ as driving frequency such that all the other elements of $\bm{\omega}$ correspond to time-dependencies in the effective Hamiltonian.

In order to construct the effective Hamiltonian explicitly, it is helpful to express the flow equation Eq.~\eqref{eq:flow} in terms of the generalized Fourier amplitudes $h_s^{\bm{m}}$.
The explicit equation of motion for the terms $h_s^{\bm{m}_0}$
and for $h_s^{\bm{m}}$ with $m_1\neq 0$ read
\begin{equation}
\begin{aligned}
\frac{dh_s^{\bm{m}_0}}{ds}&=
\frac{\bm{m}_0\cdot\bm{\omega}}{\omega_{1}}\sum_{m_1\neq 0}f_{\bm{m}}h_s^{\bm{m}}\\
&\quad\;+
\sum_{\bm{n} \atop n_1\neq 0}\frac{f_{\bm{n}}}{\omega_{1}}[h_s^{\bm{n}},h_s^{\bm{m}_0-\bm{n}}-h_s^{\bm{m}_{0}-\bm{n}_{0}}]\ ,\\
\frac{dh_s^{\bm{m}}}{ds}&=-\frac{\bm{m}\cdot\bm{\omega}}{\omega_{1}}f_{\bm{m}}h_s^{\bm{m}}+
\frac{1}{\omega_{1}}\sum_{\bm{n} \atop n_1\neq\{0,m_{1}\}}f_{\bm{m}-\bm{n}}[h_s^{\bm{m}-\bm{n}},h_s^{\bm{n}}]\\
&\quad\;+\frac{1}{\omega_{1}}\sum_{\bm{n}\atop n_1\neq 0}f_{\bm{n}}[h_s^{\bm{m}-\bm{n}_0},h_s^{\bm{n}}]\\
&\quad\;+\frac{1}{\omega_{1}}\sum_{\bm{n}}f_{\bm{m}-\bm{n}_0}[h_s^{\bm{m}-\bm{n}_0},h_s^{\bm{n}_0}]
\label{eq:dhs}
\end{aligned}
\end{equation}
with $\bm{n}_0=[0,n_2,n_3\hdots]$.
They can be solved in the well-established high-frequency expansion \cite{Bender_Orszag, PRA_Vilnius_2022} with the expansion coefficient $1/\omega_1$.

Crucially, however, the factors $\bm{m}_0\cdot\bm{\omega}/\omega_1$ and $\bm{m}\cdot\bm{\omega}/\omega_1$ do not need to be taken into account perturbatively.
While the frequency $\omega_1$ needs to be large as compared to the amplitudes in the system Hamiltonian for the high-frequency expansion to be valid, it does not need to be large as compared to the frequencies $\omega_j$ $(j>1)$.
Provided that the inequality $\bm{m}\cdot\bm{\omega}f_{\bm{m}}>0$ is satisfied for $m_1\neq 0$, the components $h_s^{\bm{m}}$ with $\bm{m}\neq\bm{m}_0$ will suffer from an exponential attenuation in the dynamics described by Eq.~\eqref{eq:dhs},
such that they vanish in the limit $s\to\infty$.
The resulting effective Hamiltonian $H_{\text{e}}$, thus only has components $h_{\text{e}}^{\bm{m}_0}=h_{s\to\infty}^{\bm{m}_0}$, {\it i.e.} time-dependence in terms of $\omega_1$ is no longer present. 


Similar to time-independent effective Hamiltonians, the high-frequency expansion of solutions of Eq.~\eqref{eq:dhs} can be specified in terms of nested commutators of  the generalized Fourier amplitudes $h^{\bm{m}}=h_{s=0}^{\bm{m}}$ of the system Hamiltonian.
The lowest order $h^{\bm{m}_{0}}_{\text{e},0}$ of the effective Hamiltonian $H_{\text{e}}$ reads
\begin{equation}
h^{\bm{m}_{0}}_{\text{e},0}=\bm{m}_{0}\cdot\bm{\omega}_{0}\sum_{m_{1}\neq0}\frac{h^{\bm{m}}}{\bm{m}\cdot\bm{\omega}}+h^{\bm{m}_{0}}\ ,
\label{first_order}
\end{equation}
and it is independent of the choice of the constants $f_{\bm{m}}$. 
The next order, that also shares this independence is specified in Eq.~\eqref{second_order} in the Appendix.

\section{Examples}

\subsection{The driven $\Lambda$-system}
A pedagogical example for the present framework is given by the realization of a time-dependent effective coupling between two low-lying states in a $\Lambda$-system with driving of two fundamental frequencies $\omega_1$ and $\omega_2$.

The system Hamiltonian reads
\begin{equation}
    H_{\Lambda}(t)=\left[\Gamma(t)+\Omega(t)e^{i\omega_{1}t}\right]|3\rangle\langle +|+\text{h.c.}\ ,
    \label{eq:HLambda}
\end{equation}
with the balanced superposition $|+\rangle=\left(|1\rangle+|2\rangle\right)/\sqrt{2}$ of two degenerate ground states,
and each of the driving functions $\Gamma(t)$ and $\Omega(t)$ is a Fourier sum with fundamental frequency $\omega_2$.

At the lowest order in the high-frequency expansion, the effective Hamiltonian reads
\begin{equation}
H_{\text{e}}^{(0)}=\sum_p\left(
\frac{p\eta}{1+p\eta}\Omega_p+\Gamma_p\right)
e^{ip\omega_2t}\ket{3}\bra{+}+\mbox{h.c.}
\label{eq:HLambda0}
\end{equation}
with the Fourier coefficients $\Omega_p$ and $\Gamma_p$ of $\Omega(t)$ and $\Gamma(t)$ and the ratio $\eta=\omega_2/\omega_1$ of the two driving frequencies.
The dependence of $H_{\text{e}}^{(0)}$ on $\eta$
reflects the fact that $\omega_1$ is not assumed to be large as compared to the second driving frequency $\omega_2$.
If this assumption was made, the leading order of the effective Hamiltonian would be independent of $\eta$ (in the quasi-static approximation), or it would contain perturbative corrections in $\eta$  (beyond the quasi-static approximation), in contrast to the actual dependence in Eq.~\eqref{eq:HLambda0}.

Resulting from the non-negligible variations in the Rabi-frequency, there is a direct coupling between the low-lying state $\ket{+}$ and the excited state $\ket{3}$, in contrast to the to regular case of the monochromatically driven $\Lambda$-system. This regular Floquet result is naturally contained in Eq.~\eqref{eq:HLambda0} in the limit $\eta\to 0$, and in the absence of any resonant driving, {\it i.e.} $\Gamma(t)=0$.

The explicit dependence of $H_{\text{e}}^{(0)}$ on the driving parameters can also be used to identify driving profiles that ensure that no undesired excitations to the excited state $\ket{3}$ occur.
For any component $\Omega_m$ of the off-resonant drive, the corresponding component $\Gamma_m$ of the resonant drive can be chosen such that $H_{\text{e}}^{(0)}$ vanishes.

\begin{figure}
\includegraphics[width=0.45\textwidth]{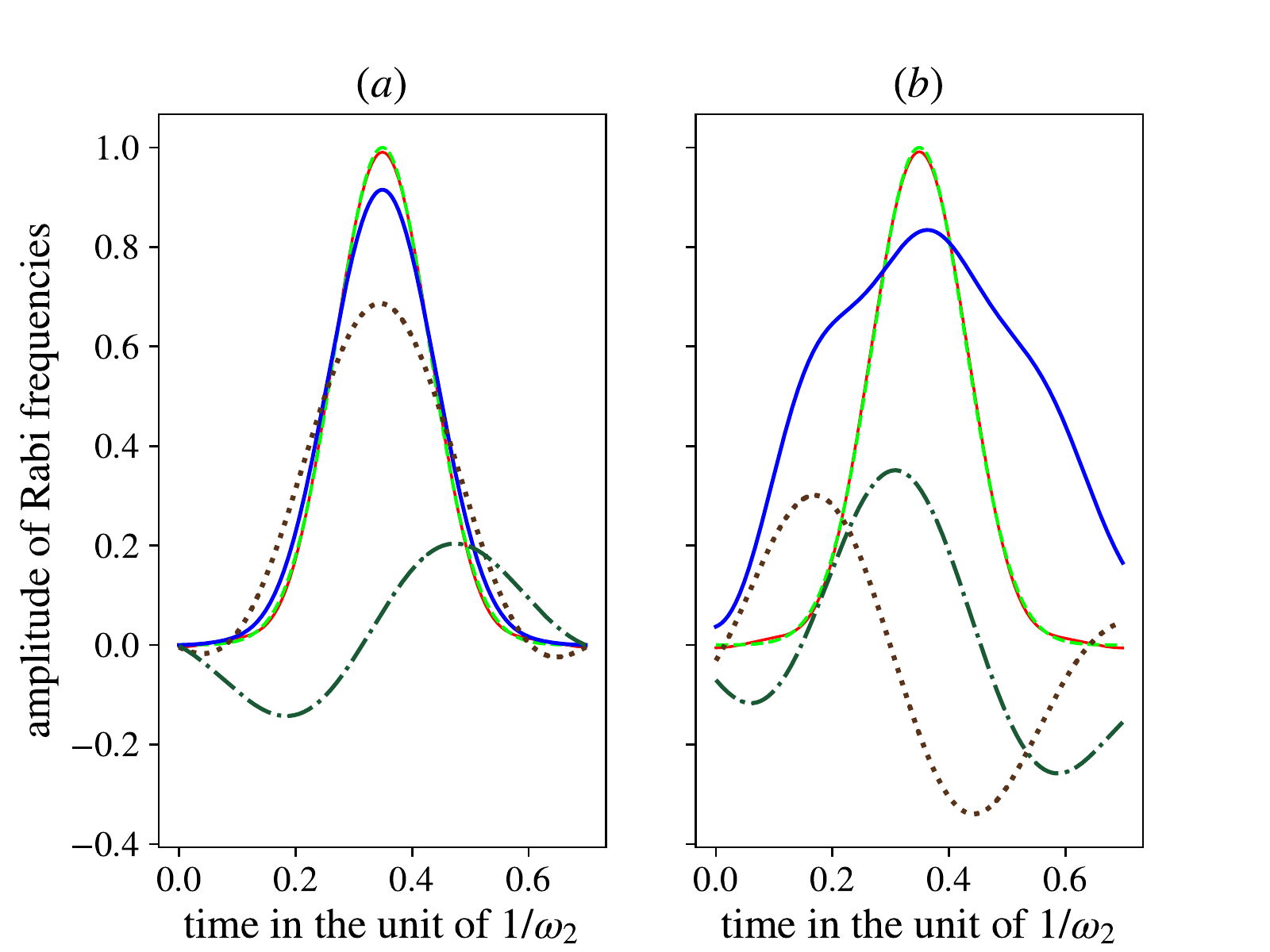}
\caption{Time-dependent effective Rabi-frequency $\Omega_e(t)$ of the effective Hamiltonian $H_{\text{e}}^{(1)}(t)$ and underlying driving profile $\Omega(t)$.
Insets a) and b)correspond to parameter values $\eta=1/(5\sqrt{3})$ and $\eta=1/\sqrt{7}$ respectively.
The thin dashed green line depicts a normalized Gaussian profile as targeted time-dependence.
The solid red line depicts the actual effective Rabi-frequency (normalized) following Eq.~\eqref{eq:cm} realized with $7$ Fourier components (between $-3$ and $3$);
the solid blue line depicts the quasi-static approximation ($\eta\to 0$) of the effective Rabi-frequency.
The dotted dark red and dashed dotted dark green lines depict the real and imaginary part of the driving profile $\Omega(t)$.}
\label{Fig:Lambda}
\end{figure}

Given this choice, the first order contribution to the effective Hamiltonian reads
\be
H_{\text{e}}^{(1)}(t)=\Omega_{\text{e}}(t)\bigl(|3\rangle\langle 3|-|+\rangle\langle+|\bigr)\ ,
\ee
with the effective Rabi-frequency
\be
\Omega_{\text{e}}(t)=\sum_{p,q}
\frac{\left[1+(2p-q)\eta\right]\exp(iq\omega_2t)}{(1-q^2\eta^2)(1+p\eta)\left[1+(p-q)\eta\right]}\frac{\Omega_p\Omega_{p-q}^\ast}{\omega_1}\ .
\label{eq:cm}
\ee
In the limit $\eta\to 0$, the effective Rabi frequency reduces to
\be
\Omega_{\text{qs}}(t)=\sum_{p,q}
\Omega_p\Omega_{p-q}^\ast\frac{\exp(iq\omega_2t)}{\omega_1}=\frac{|\Omega(t)|^2}{\omega_1}\ ,
\ee
{\it i.e.} to the quasi-static solution.

Eq.~\eqref{eq:cm} permits to identify driving profiles $\Omega(t)$ that realize a desired time-dependent effective Rabi frequency $\Omega_{\text{e}}(t)$.
Fig.~\ref{Fig:Lambda} depicts the exemplary case of a Gaussian time-dependence for the effective coupling between the two low-lying states.
Insets $(a)$ and $(b)$ correspond to the parameter values $\eta=1/(5\sqrt{3})$ and $\eta=1/\sqrt{7}$ {\it i.e.} in one case the adiabatic approximation is expected to hold approximately while in the other the adiabatic approximation to be violated.
The thin dashed green line depicts the desired time-dependence; the actual effective Rabi-frequency $\Omega_{\text{e}}(t)$ can be made to approximate the desired time-dependence arbitrarily well, but a restriction to a finite number of $7$ Fourier components $\Omega_{m}$ results in a small deviation from the desired behavior.

The solid orange-red line depicts the quasi-static approximation $\Omega_{\text{qs}}(t)$; in inset $(a)$, this approximation is indeed good, but inset $(b)$ shows that a clear separation of time-scales ({\it i.e.} $\eta\ll 1$) is required for the quasi-static approximation to hold.

The dotted dark red and dashed dotted dark green lines depict the real and imaginary parts of the actual driving function $\Omega(t)$ in Eq.~\eqref{eq:HLambda}.
Given the validity of the quasi-static approximation (Fig.~\ref{Fig:Lambda}(a)), the driving function $\Omega(t)$ is approximately mirror symmetric/anti-symmetric around the mid-point of the depicted time-window,
and in the quasi-static limit, this symmetry of the targeted Gaussian dependence is given exactly.
Outside the regime of validity of the quasi-static approximation (Fig.~\ref{Fig:Lambda}(b)) this symmetry is clearly violated by $\Omega(t)$, highlighting that a diabatic increase of $\Omega(t)$ requires different driving than a diabatic decrease.

\subsection{Quench across phase-transitions of a Chern insulator}
\label{sec:quench}

\begin{figure}
\includegraphics[width=0.45\textwidth]{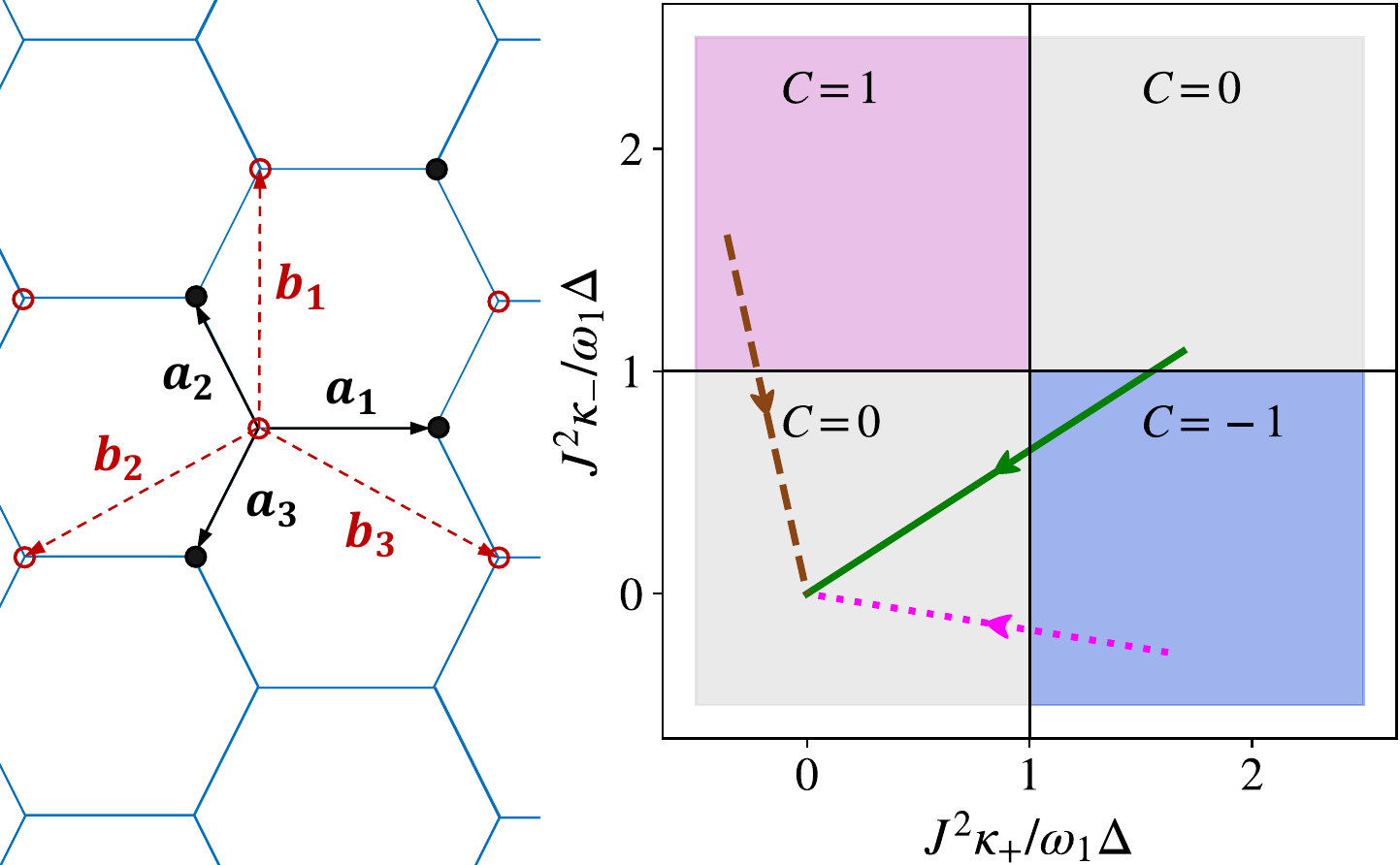}
\caption{Geometry of a honeycomb lattice (left) and quenches through a phase diagram (right). The honeycomb lattice is comprised of two triangular sub-lattices, consist of black and hollow red dots. Directions of nearest neighbour and next nearest neighbour tunnelling processes are depicted with black and dashed red vectors. An exponential decrease of the tunnelling rate $J(t)$ satisfying the conditions Eq.~\eqref{eq:fine_tune} allows exploration the two-dimensional phase diagram spanned by $J^{2}\kappa_{+}/\omega_{1}\Delta$ and $J^{2}\kappa_{-}/\omega_{1}\Delta$. Grey regions corresponds to Chern number being $0$, while purple and blue regions correspond to $C=\pm 1$. Three quenches profiles in dashed lines are plotted based on three solutions of Eq.~\eqref{eq:fine_tune} with the sequence $\{q_{k},\omega_{k},\delta_{k}-\delta_{k}^{\prime}\}$ inherited from~\cite{tunable_Chern}.}
\label{Fig:Lattice}
\end{figure}

A more involved example is given by the problem of crossing of a phase transition in a Chern insulator \cite{Kun_Yang,Cooper_Quench_Chern}.
It is defined by a driven non-interacting Hubbard Hamiltonian $H=H_\text{B}+H_\text{S}$ with
\be
H_\text{B}=-J(t)\sum_{\langle i,j\rangle}c_i^\dagger c_j+\sum_i\Delta_ic_i^\dagger c_i,\,\hspace{.2cm} 
H_\text{S}=\sum_{i}V_i(t)c^{\dagger}_{i}c_{i}\ ;
\label{eq:HBH}
\ee
$c_j^\dagger$ and $c_i$ are creation and annihilation operators on a hexagonal lattice and $\langle...\rangle$ denotes the nearest neighbours. This lattice is given by a triangular Bravais lattice and two-site basis, or, equivalently by two triangular sub-lattices, depicted in Fig.~\ref{Fig:Lattice} by red empty (sub-lattice $A$) and black full (sub-lattice $B$) circles respectively. The onsite energies $\Delta_i$ are chosen such that all sites on sub-lattice $A$ have the same onsite-energy $\Delta$ and all sites on sub-lattice $B$ have the opposite onsite-energy $-\Delta$. There are three inequivalent directions of nearest neighbor tunnelling processes depicted by $\bm{a}_1$, $\bm{a}_2$ and  $\bm{a}_3$. The tunnelling rate $J(t)$ for those processes does not depend on the direction in real space, but it is time-dependent. All tunnelling processes beyond nearest neighbors are neglected. The Hamiltonian $H_\text{S}$ in Eq.~\eqref{eq:HBH}
describes shaking~\cite{Nature_Esslinger_Haldane, Hanns_and_Florian, Quench_Haldane_Experiment, US_Shaken_1D_Lattice}
in terms of time-dependent onsite energies
\begin{equation}
V_i(t)=\sum_{k=1}^{N}q_{k}\omega_{k}\left[\cos\left(\omega_{k}t-\delta_{k}\right)x_i+\cos\left(\omega_{k}t-\delta^{\prime}_{k}\right)y_i\right]\,
\label{eq:shaking}
\end{equation}
with driving amplitudes $q_k$, driving frequencies $\omega_k$ and phases $\delta_k$ and $\delta_k^\prime$~\cite{tunable_Chern}, and the positions $[x_i,y_i]$ of the lattice sites.
In the frame defined by the shaking term $H_\text{S}$, the resulting effective Hamiltonian has the same type of tunnelling processes as $H_\text{B}$ in Eq.~\eqref{eq:HBH},
but the tunnelling rates are renormalized, and they are generally complex. For suitable phase relations between the different tunnelling processes, the effective Hamiltonian captures the Haldane Model~\cite{Haldane, Nature_Esslinger_Haldane} with two topological non-trivial phases (with Chern number $+1$ and $-1$) and one topologically trivial phase (with Chern number $0$). Deviations from these phase relations result in a deformation of the system's phase diagram~\cite{tunable_Chern} (typically depicted in terms on onsite energy and phase of next-nearest-neighbor tunnelling rate), but, they do not affect the existence of several phases with different topological properties.

An effective Hamiltonian with time-dependent onsite energy can be used to investigate the creation of topological defect generations~\cite{Kun_Yang} as the system is quenched through the phase boundary between topological trivial and non-trivial phases.
In addition to the effective onsite energies $\tilde\Delta(t)$ and $-\tilde{\Delta}(t)$ for sub-lattices $A$ and $B$, 
the effective Hamiltonian has nearest-neighbor tunnelling rates $\tilde J_k(t)$  (from sub-lattices $A$ to $B$ along ${\bf a}_k$ ($k=1,2,3$) in Fig.~\ref{Fig:Lattice}),
and the rates $\tilde G_k(t)$ ($-\tilde{G}_{k}(t)$) of next-nearest neighbor tunnelling processes (along ${\bf b}_k$ ($k=1,2,3$) in Fig.~\ref{Fig:Lattice}) within sublattice $A$ ($B$).
All these quantities can be obtained without assuming the separation of time-scales discussed above, but for the sake of clarity, the following discussion is focused on the dominant corrections to the quasi-static approximation.

The effective Hamiltonian in the frame induced  by the shaking Hamiltonian $H_\text{S}(t)$ is characterized by the system parameters
\begin{equation}
\begin{aligned}
    \tilde\Delta(t)&=\Delta-\frac{1}{\omega_1}\left({J^{2}(t)}C_{\Delta}+\dot{J}(t)J(t)\tilde{C}_{\Delta}\right)\ ,\\
    \tilde J_{k}(t)&=J(t)D_{k}-\frac{1}{\omega_{1}}\dot{J}(t)\tilde{D}_{k}\ ,\\
    \tilde G_{k}(t)&=-\frac{1}{\omega_{1}}\left(J^{2}(t)E_{k}+\dot{J}(t)J(t)\tilde{E}_{k}
    \right)\ ,\label{parameters}
\end{aligned}
\end{equation}
where $C_{\Delta}$, $\tilde{C}_{\Delta}$, $D_{k}$, $\tilde{D}_{k}$, $E_{k}$ and $\tilde{E}_{k}$ (given in Eq.~\eqref{Coe} and Eq.~\eqref{v_and_p}) are time-independent scalars that are specific to the hexagonal lattice geometry and that depend on the amplitudes $q_k$, frequencies $\omega_k$ and phases $\delta_k$, $\delta_k^\prime$ of the shaking profile in Eq.~\eqref{eq:shaking}.
 With a suitable time-dependence of the tunnelling rate $J$ and the shaking profile, one can realize a broad range of time-dependencies in the effective Hamiltonian, as exemplified in the following.

The lattice of the underlying Hamiltonian (Eq.~\eqref{eq:HBH}), and many lattice models of interest such as the Haldane model or the Kitaev model~\cite{KITAEV2006}, are invariant under a rotation of $2\pi/3$.
The shaking necessarily breaks this invariance so that the resulting effective Hamiltonian does not satisfy this symmetry for general driving patterns.
The symmetry can, however, be partially recovered with suitably chosen shaking profiles.
For the given time-dependent tunnelling rate $J(t)=J_{0}e^{-\gamma t}$ for example,
the condition
\begin{equation}
D_{1}+\frac{\gamma}{\omega_{1}}\tilde{D}_{1}=D_{2}+\frac{\gamma}{\omega_{1}}\tilde{D}_{2}=D_{3}+\frac{\gamma}{\omega_{1}}\tilde{D}_{3}
\label{eq:fine_tune}
\end{equation}
ensures the desired symmetry for the nearest neighbor tunnelling, {\it i.e.} $\tilde J_1=\tilde J_2=\tilde J_3$ (though next-to-nearest tunnelings are still anisotropic), and this condition can indeed be satisfied for tri-chromatic shaking profile, {\it i.e.} $N=3$ in Eq.~\eqref{eq:shaking}.
The Chern number of the resultant system is given by $C=\frac{1}{2}\left[\mathrm{sgn}(h_{+})-\mathrm{sgn}(h_{-})\right]$~\cite{analytical_Chern} with
$h_{\pm}=\tilde\Delta+2\sum_{k}|\tilde G_{k}|\cos\left(\alpha_{k}\pm2\pi/3\right)$,
where $\alpha_{k}$ ($k=1,2,3$) are the phases of complex NNN hopping rates \footnote{The Fourier transform convention on the fermionic creation and annihilation operators is $\bm{c}_{\bm{r}_{i}}=\frac{1}{\sqrt{N}}\sum_{\bm{k}}\bm{c}_{\bm{k}}e^{i\bm{k}\cdot\bm{r}_{i}}$, where $\bm{r}_{i}$ denotes the center of unit cells, $\bm{c}_{\bm{r}_{i}}$ is the spinor on the two-site basis of the unit cell located at $\bm{r}_{i}$, and $N$ is the number of unit cells.}. The parameters $h_{\pm}$ that the Chern number depends on can also be expressed as $h_{\pm}=\Delta-J^{2}(t) \kappa_{\pm}/\omega_1$, where
$\kappa_{\pm}=\tau_{1}+2\sum_{k}\tilde{\tau}_{k}\cos\left(\alpha_{k}\pm2\pi/3\right)$ with $\tau_{1}$ and $\tilde{\tau}_{k}$ given in Eq. ~\eqref{taus} in the Supp. Mat. are time-independent.
The system can thus be characterized in terms of a phase diagram spanned by $J^{2}\kappa_+/\omega_{1}\Delta$ and $J^{2}\kappa_-/\omega_{1}\Delta$, as depicted in the right panel of Fig.~\ref{Fig:Lattice}.

Also shown are three exemplary solutions for quenches with exponential time-dependence that can be realised in terms of suitably modulated lattice shaking.
The quenches depicted by dashed brown and dotted magenta lines correspond to initial conditions in topologically non-trivial phases, and a final point in a topologically trivial phase, whereas the start and the end point of the quench depicted with a solid green line lies in domains of topologically trivial phase, but the quench takes the system through an ordered phase. In all these cases, the rate $\gamma$ in the tunnelling rate $J(t)$ can be varied from a regime of adiabatic to diabatic quenches.

\section{Range of applicability}

In order to gauge the range of applicability of the involved approximations, this section provides a comparison with numerically exact simulations utilising the \texttt{QuSpin} package \cite{Quspin_1, Quspin_2}.
 
\subsection{Frequency regime}
\label{sec:validity}

\begin{figure*}[b]
\centering
\begin{subfigure}{.32\textwidth}
\centering
\includegraphics[width=1\textwidth]{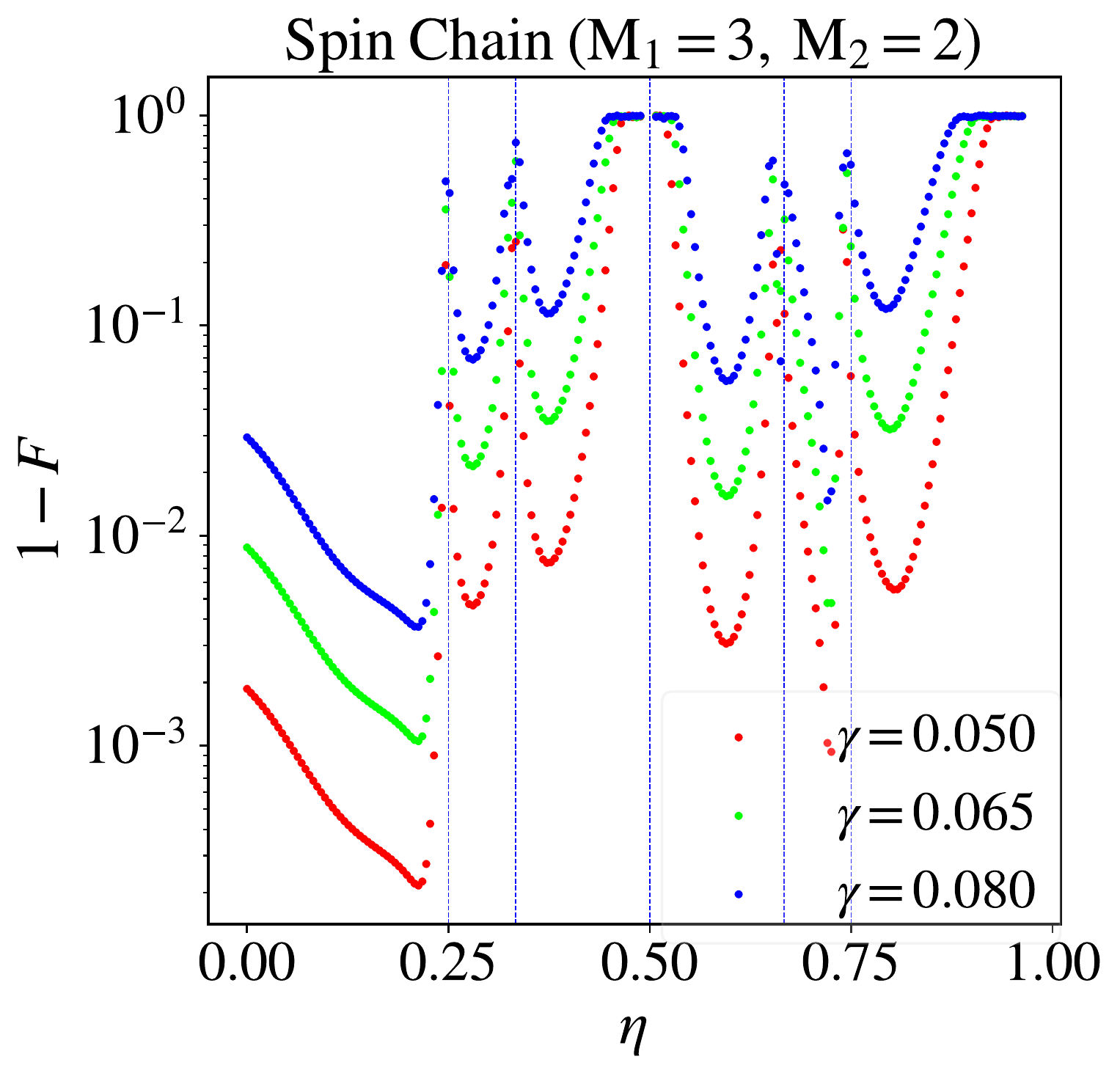}
\label{fig:spin_chain_infidelity_3_2}
\caption*{$(a)$}
\end{subfigure}
\begin{subfigure}{.32\textwidth}
\centering
\includegraphics[width=\textwidth]{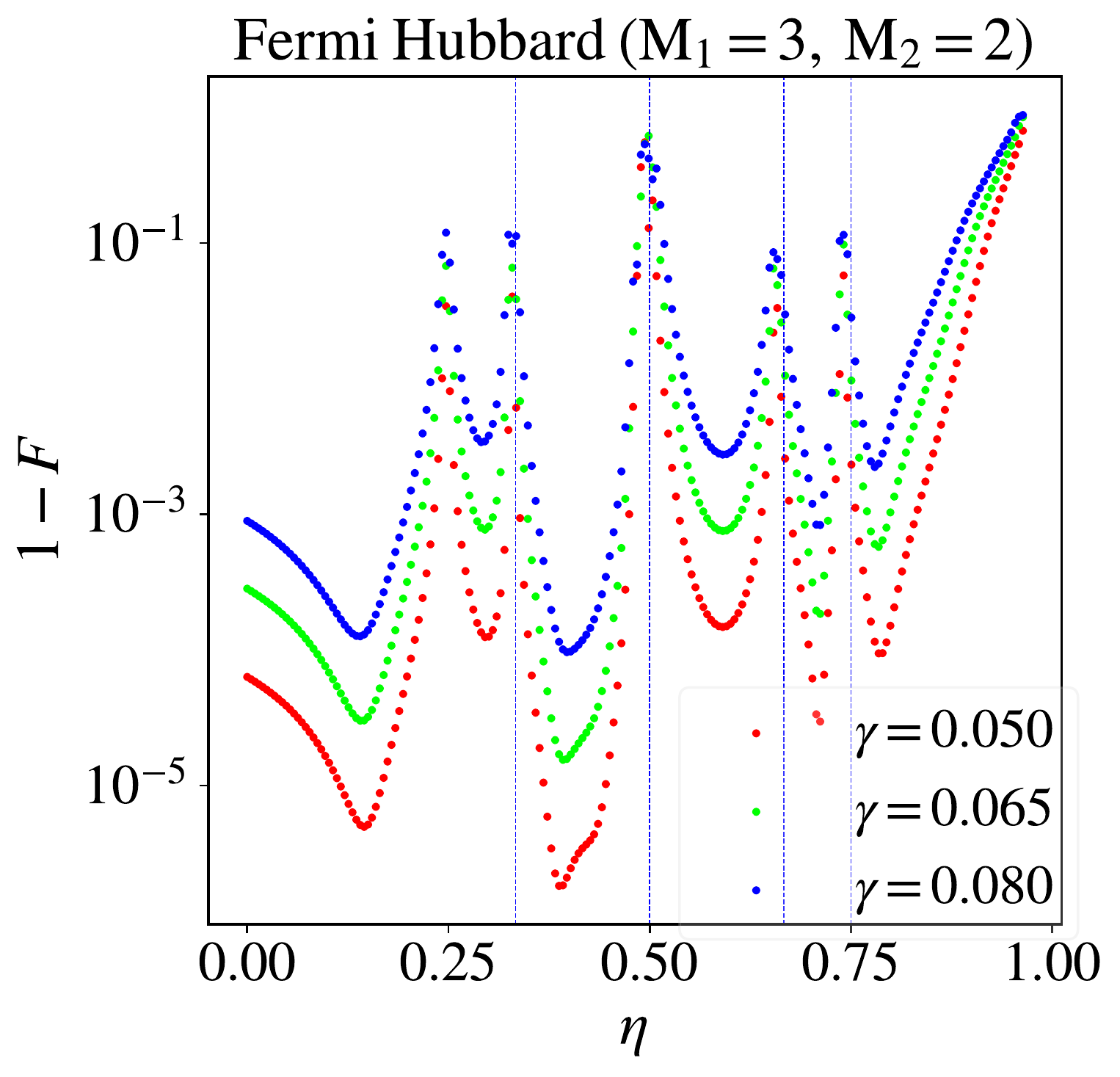}
\label{fig:fermi_hubbard_infidelity_3_2}
\caption*{$(b)$}
\end{subfigure}
\begin{subfigure}{.32\textwidth}
\centering
\includegraphics[width=\textwidth]{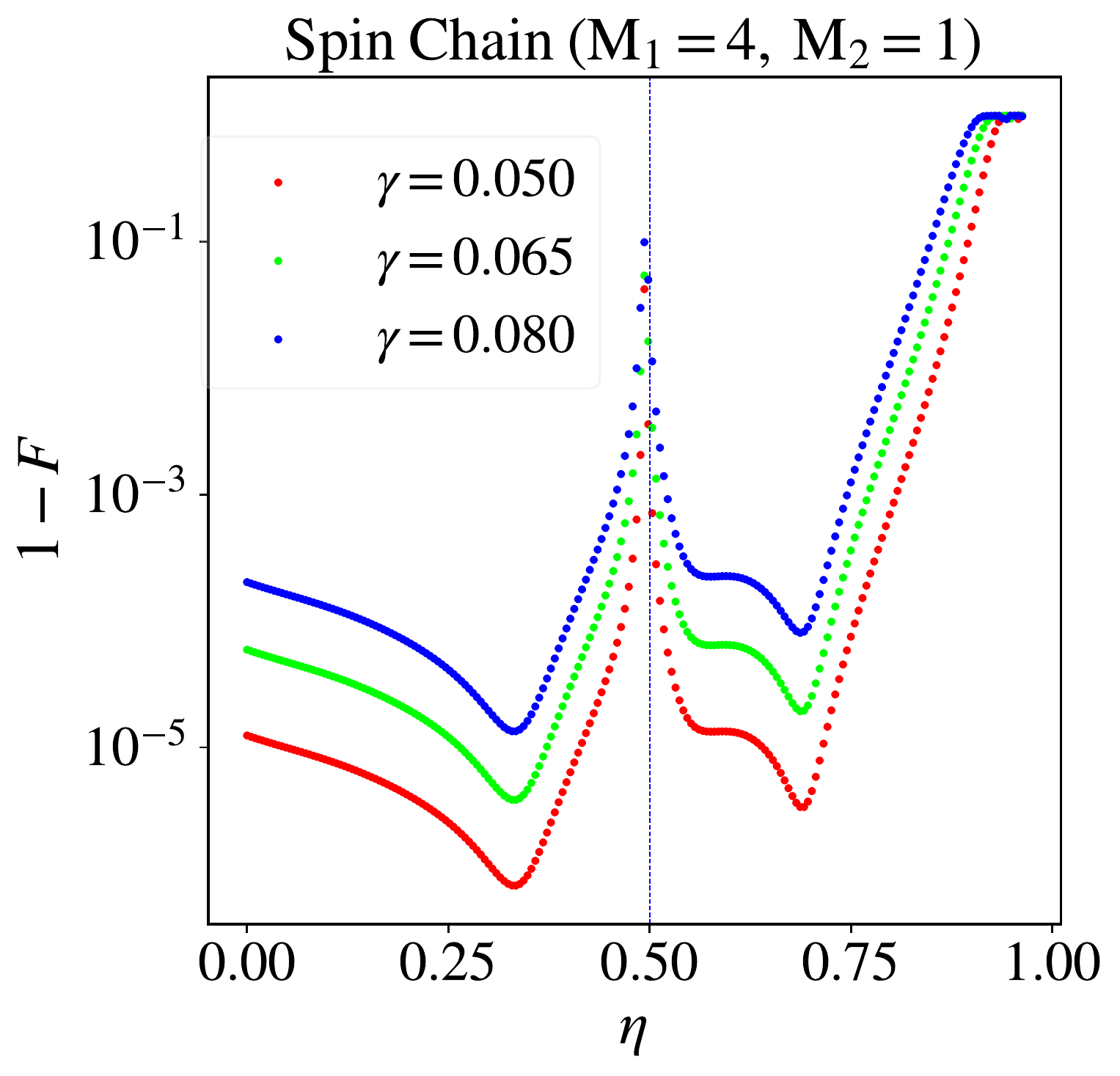}
\label{fig:spin_chain_infidelity_4_1}
\caption*{$(c)$}
\end{subfigure}
\caption{Gate infidelity (Eq.~\eqref{eq:gatefidelity}) for the driven spin-chain, (a) and (c), and Fermi-Hubbard model (b) as a function of $\eta=\omega_2/\omega_1$ for different values of $\gamma$ (defined in Eq.~\eqref{A_Fourier}).
The dynamics in the Fermi-Hubbard model takes place in the zero-quasi-momentum subspace of $5$ fermions with $2$ spin-up and $3$ spin-down.
Dashed blue lines indicate the rational values $1/4$, $1/3$, $1/2$, $2/3$, $3/4$ for $\eta$ around which the multi-mode Floquet theory is expected to break down.}
\label{fig:fidelity}
\end{figure*}

The present expansion is derived under the premise of the high-frequency expansion ({\it i.e.} $\omega_1$ exceeds all relevant rates in the Hamiltonian and the inequalities $\omega_1\gtrsim\omega_j$.
Since the time-dependence in the underlying Hamiltonian $H(t)$ can include higher harmonics of the frequencies $\omega_j$, this implies that there can be frequency components in the time-dependent effective Hamiltonian that exceed the fundamental driving frequency $\omega_1$.
This is corroborated in Fig.~\ref{fig:fidelity} that depicts the infidelity of the dynamics induced by the time-dependent effective Hamiltonian for a spin chain comprised of $16$ interacting spins and for a one-dimensional Fermi-Hubbard system with $16$ sites.

The underlying Hamiltonian for the spin chain reads
\begin{equation}
    H(t)=d_{x}(t)\sum_{i}\sigma_{x}^{i}+d_{zz}(t)\sum_{i}\sigma_{z}^{i}\sigma_{z}^{i+1}+d_{yy}(t)\sum_{i}\sigma_{y}^{i}\sigma_{y}^{i+1},
\label{eq:spin_chain_H}
\end{equation}
with the time-dependent functions $d_{x}(t)$, $d_{zz}(t)$ and $d_{yy}(t)$.
The underlying Hamiltonian for the Fermi-Hubbard chain with $L$ sites reads
\begin{equation}
\begin{aligned}
H(t)&=-J(t)\sum_{i=0}^{L-1}\sum_{\sigma=\downarrow,\uparrow}\left(c^{\dagger}_{i+1,\sigma}c_{i,\sigma}+\mathrm{h.c.}\right)\\
&\quad+\;U_{I}(t)\sum_{i=0}^{L-1}c_{i,\uparrow}^\dagger c_{i,\uparrow} c_{i,\downarrow}^\dagger c_{i,\downarrow}\ ,
\end{aligned}
\label{eq:fermi_hubbard_H}
\end{equation}
with annihilation operator $c_{i,\sigma}$ of a Fermion of spin $\sigma$ at site $i$ and corresponding creation operator $c_{i,\sigma}^\dagger$,
and with time-dependent tunnelling amplitude $J(t)$ and interaction rate $U_{I}(t)$.
Both models are understood with periodic boundary conditions. The resulting translational invariance implies conservation of quasi-momentum, and the following discussion is focused on the subspace with zero quasi-momentum.

With the propagator $U_{0}(t)$ induced by the underlying Hamiltonian and the propagator $U_\text{e}(t)$ induced by the time-dependent effective Hamiltonian in first order in $1/\omega_1$,
the fidelity of the effective dynamics at $t=2\pi/\omega_1$ is defined as \cite{fidelity_definition}
\begin{equation}
    F(U_{0}, U_e)=\mathrm{Re}\left[\mathrm{Tr}(U^{\dagger}_{0}U_e)\right]/\mathrm{dim}(\mathcal{H})\ ,
    \label{eq:gatefidelity}
\end{equation}
where the factor $\mathrm{dim}(\mathcal{H})$ in terms of the dimension of the Hilbert space ensures a maximal value of $1$ of the fidelity.
Fig.~\ref{fig:fidelity} depicts the infidelity $1-F$ (in log-scale) as a function of $\eta=\omega_2/\omega_1$.

The data shown in Fig.~\ref{fig:fidelity}, is based on random choices of the driving functions $d_{x}(t)$, $d_{zz}(t)$, $d_{yy}(t)$, $J(t)$ and $U_{I}(t)$ in terms of bi-chromatic Fourier sums
\begin{equation}
\gamma\,\omega_{1}\sum_{p=-M_1}^{M_1}\sum_{q=-M_2}^{M_2}A_{pq}e^{i(p\omega_{1}+q\omega_{2})t}\ ,\label{A_Fourier}
\end{equation}
where the real parameter $\gamma$ 
is the ratio between driving strength and driving frequency $\omega_1$; {\it i.e.} the high-frequency approximation is valid for $\gamma\ll 1$.
The integers $M_1$ and $M_2$ specify the spectral width of the driving functions, and the complex numbers $A_{pq}$ are chosen at random from within the interval $[-0.5, 0.5]$.
All the driving parameters are available on \cite{zenodo}.

The left-most data points (for $\eta=0$) in each of the three panels  correspond to the regular Floquet case with a time-independent effective Hamiltonian,
and the infidelities obtained for $\eta=0$ give an idea of what infidelity one can reasonably expect for a given value of $\gamma$.
In large parts of the parameter space, the infidelities obtained for the time-dependent effective Hamiltonian are comparable or even lower than in the regular Floquet case.
Only when $\eta$ is approximated well by a rational number $p/q$ with small integers $p$ and $q$, there is a sizeable increase of the infidelity, {\it i.e.} a decrease in fidelity, as expected from the breakdown of multi-mode Floquet theory that requires incommensurate frequencies. The increase in infidelity is most pronounced for $\eta\simeq 1/2$ and for $\eta\simeq 1$, but also visible for $\eta\simeq 1/4$, $\eta\simeq 1/3$, $\eta\simeq 2/3$ and $\eta\simeq 3/4$ in insets $(a)$ and $(b)$.
Inset $(c)$ shows fewer instances of increased infidelity, since it is based on driving patterns with fewer $\omega_{2}$ frequency components ($M_1=4$ and $M_2=1$ as opposed to $M_1=3$ and $M_2=2$ in insets $(a)$ and $(b)$),
highlighting that the accuracy of effective Hamiltonians at given values of $\eta$ can be controlled through the spectral properties of the driving functions and a careful design of driving functions is essential for a challenging quantum simulation such as the quench dynamics discussed in Sec.~\ref{sec:quench}.

\subsection{Long-time dynamics and heating}
\label{sec:heating}

A crucial issue with driven quantum systems is heating, and as shown in the following the heating resultant from the present driving schemes is comparable to the heating obtained in regular Floquet engineering of time-independent effective Hamiltonians.

\begin{figure*}[b]
\centering
\begin{subfigure}{.463\textwidth}
\centering
\includegraphics[width=\textwidth]{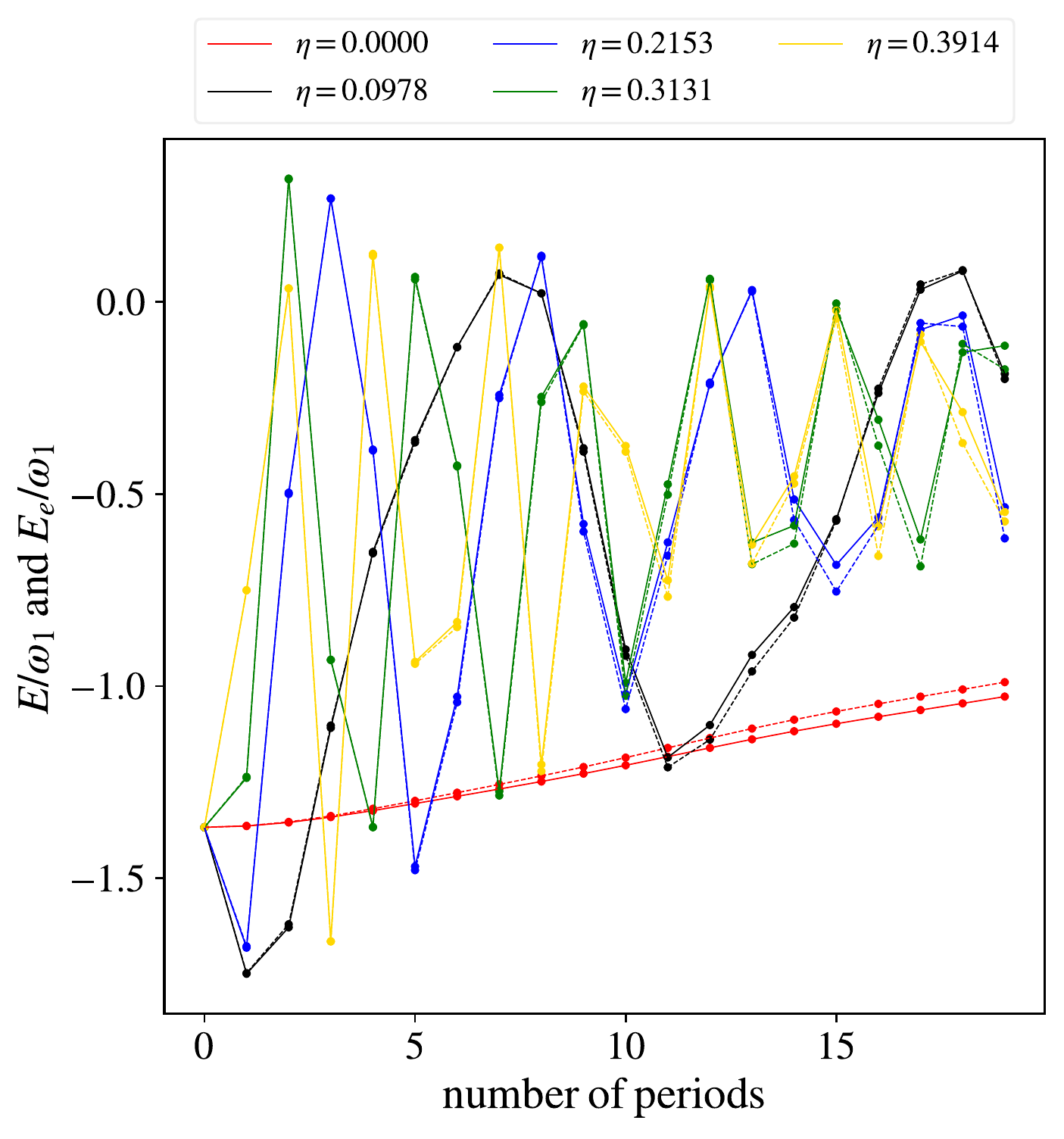}
\label{fig:fermi_hubbard_energy}
\caption*{$(a)$}
\end{subfigure}
\begin{subfigure}{.463\textwidth}
\centering
\includegraphics[width=\textwidth]{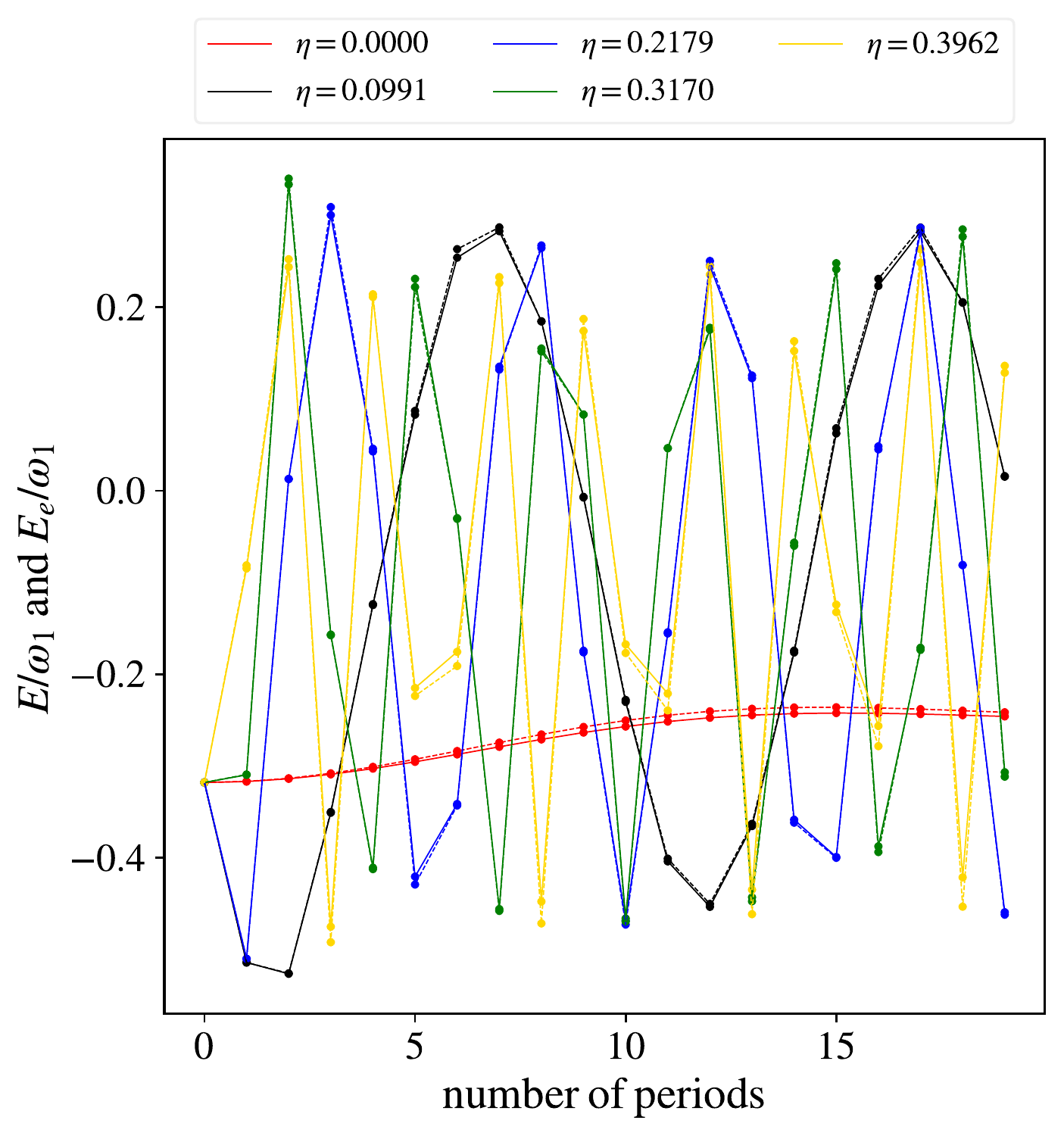}
\label{fig:fermi_hubbard_energy}
\caption*{$(b)$}
\end{subfigure}
\caption{Stroboscopic dynamics of energy expectation of the driven spin-chain system with $16$ sites (left) and driven Fermi-Hubbard model (right) with $16$ sites with the same filling information as in Fig. \ref{fig:fidelity} up to the $19\times 2\pi/\omega_{1}$.
The dots represent numerical data while lines connecting them should help to distinguish between the dynamics of the driven system (solid) and effective dynamics (dashed).} 
\label{fig:heating}
\end{figure*}

Heating is most suitable characterized in term of a comparison between the energy expectations of the driven system and the quantum-simulated system over several driving periods.
The expectation value of the driven Hamiltonian for both of these dynamics is shown in Fig.~\ref{fig:heating} for the spin chain and the Fermi-Hubbard chain (with the same system parameters as above in Sec.~\ref{sec:validity},
and with the system initialized in ground state of the bare Hamiltonian for $\eta=0$ at $t=0$ and $\gamma=0.06)$.
The spectral width of the driving functions are characterized by $M_{1}=3$ and $M_{2}=1$ for both the spin chain and the Fermi-Hubbard model.

Energy expectation values are depicted by dots (with different colors for different values of $\eta$.
The dots are connected with straight lines to guide the eye and to distinguish between energy expectations of the driven system (solid) and the quantum-simulated system (dashed).

In both insets one can see that the energy difference between the driven and quantum-simulated systems are small as compared to the energy expectations,
and that these differences to not grow noticeably with time or with $\eta$. This suggests that the heating caused by polychromatic driving is sufficiently low over several driving periods that is does not jeopardize an accurate quantum simulation.

\section{Conclusions}
While the field of quantum simulations of time-independent quantum systems has demonstrated the readiness of quantum mechanical hardware for problems that are far outside the range of classically achievable simulations,
ideas for quantum simulations of time-dependent Hamiltonians~\cite{PRA_Vilnius_2022} are only in its infancy.
The ability to realize time-dependencies that do not need to be slow as compared to the driving time-scale and that can be designed with several fundamental frequencies enables the experimental realization of quantum simulations of a broad range of physical problems with explicit time-dependence in numerous state of the art platforms, such as sweeping between different quantum phases in the lattice $\mathbb{Z}_{2}$ gauge theories~\cite{Monika_2019_Z_2} and the Hofstadter model~\cite{Dalibard_PRX} or realizing time-dependent Hamiltonians that exhibits Floquet symmetry-protected topological phases~\cite{SPT, SPT_exp}.

\begin{acknowledgments}
We gratefully acknowledge stimulating discussions with Florian Meinert, Hanns-Christoph N\"agerl, Johannes Knolle and Viktor Novičenko. Numerical simulations were performed on the Imperial HPC cluster with the QuSpin package~\cite{Quspin_1, Quspin_2}. We thank the Imperial HPC team for providing valuable support for our work. Boyuan Shi is supported by the Imperial College President's Scholarship.
\end{acknowledgments}
\clearpage
\newpage
\appendix 
\makeatletter
\patchcmd{\@makechapterhead}{\thechapter}{{\Huge Appendix }\thechapter}{}{}
\makeatother
\section{High-Frequency Expansions of the flow equation up to $\mathcal{O}\left(1/\omega_{1}\right)$ Order}

The first-order component of the effective Hamiltonian is
\begin{widetext}
\begin{equation}
\begin{aligned}
&\frac{1}{\omega_{1}}h^{\bm{m}_{0}}_{e,1}=-\sum_{n_{1}\neq0,\bm{n}_{0}}\frac{n_{1}\omega_{1}}{\left[\bm{n}_{0}\cdot\bm{\omega}_{0}+n_{1}\omega_{1}\right](\bm{m}_{0}\cdot\bm{\omega}_{0}+n_{1}\omega_{1})}[h^{\bm{n}},h^{\bm{m}_{0}-\bm{n}_{0}}]\\
&-\sum_{n_{1}\neq0,m_{1}\neq0,\bm{n}_{0}}\frac{m_{1}\omega_{1}\,(\bm{m}_{0}-\bm{n}_{0})\cdot\bm{\omega}_{0}+(\bm{m}_{0}\cdot\bm{\omega}_{0})^{2}}{\left[(\bm{m}_{0}-\bm{n}_{0})\cdot\bm{\omega}_{0}+m_{1}\omega_{1}\right]\bm{n}\cdot\bm{\omega}\,\bm{m}\cdot\bm{\omega}}[h^{\bm{n}},h^{(m_{1},\bm{m}_{0}-\bm{n}_{0})}]\\
&-\frac{1}{2}\sum_{n_{1}\neq0,\bm{n}_{0}}\frac{n_{1}^{2}\omega_{1}^{2}\left[(\bm{m}_{0}-2\bm{n}_{0})\cdot\bm{\omega}_{0}-n_{1}\omega_{1}\right]}{ \bm{n}\cdot\bm{\omega}\left[(\bm{m}_{0}\cdot\bm{\omega}_{0})^{2}-n_{1}^{2}\omega_{1}^{2}\right]\left[(\bm{m}_{0}-\bm{n}_{0})\cdot\bm{\omega}_{0}-n_{1}\omega_{1}\right]}[h^{\bm{n}},h^{(-n_{1},\bm{m}_{0}-\bm{n}_{0})}]\\
&-\frac{1}{2}\sum_{n_{1}\neq\{0,m_{1}\},\atop m_{1}\neq0,\bm{n}_{0}}
(m_{1}-n_{1})n_1\omega_{1}^2
\frac{\bm{m}_{0}\cdot\bm{\omega}_{0}
((2\bm{n}_{0}-\bm{m}_{0})\cdot\bm{\omega}_{0}-(m_{1}-n_{1})\omega_{1})+
m_{1}\omega_{1}\bm{n}_{0}\cdot\bm{\omega}_{0}}
{\bm{m}\cdot\bm{\omega}\,\bm{n}\cdot\bm{\omega}\,(\bm{m}-\bm{n})\cdot\bm{\omega}\,(\bm{m}_{0}\cdot\bm{\omega}_{0}+n_{1}\omega_{1})[\bm{m}_{0}\cdot\bm{\omega}_{0}+(m_{1}-n_{1})\omega_{1}]}[h^{\bm{n}}, h^{\bm{m}-\bm{n}}].
\label{second_order}
\end{aligned}
\end{equation}
\end{widetext}

If we treat $|\bm{\omega}_{0}|/\omega_{1}$ as the same order as $|H(t)|/\omega_{1}$, then Eq. \eqref{second_order} turns to 
\begin{equation}
\begin{aligned}
\frac{1}{\omega_{1}}h_{e,1}^{\bm{m}_{0}}&=\frac{1}{\omega_{1}}\sum_{n_{1}\neq0,\bm{n}_{0}}\frac{1}{n_{1}}[h^{\bm{n}},h^{\bm{m}_{0}-\bm{n}_{0}}]\\
&\quad+\frac{1}{2\omega_{1}}\sum_{n_{1}\neq0,\bm{n}_{0}}\frac{1}{n_{1}}[h^{\bm{n}},h^{(-n_{1},\bm{m}_{0}-\bm{n}_{0})}],
\end{aligned}
\end{equation}
which along with the zeroth-order expression in the same limit,
\begin{equation}
h_{e,0}^{\bm{m}_{0}}=\bm{m}_{0}\cdot\bm{\omega}_{0}\sum_{m_{1}\neq0}\frac{h^{\bm{m}}}{n_{1}\omega_{1}}+h^{\bm{m}_{0}}
\end{equation}
would be identical to the results using the framework in \cite{PRA_Vilnius_2022} in the Floquet gauge by further decomposing $h^{(n_{1})}(t)=\sum_{\bm{m}_{0}}h^{(n_{1},\bm{m}_{0})}e^{i\bm{m}_{0}\cdot\bm{\omega}_{0}t}$.
This shows that the flow equation in the main text is consistent with the flow equation constructed in \cite{PRA_Vilnius_2022} under the same limit.
\section{Explicit Expressions of $C_{\Delta}$, $\tilde{C}_{\Delta}$, $D_{k}$, $\tilde{D}_{k}$, $E_{k}$, $\tilde{E}_{k}$, $\tau_{1}$ and $\tilde{\tau}_{k}$}
\label{Cs}

Taylor expanding Eq. \eqref{first_order} and Eq. \eqref{second_order} to the first order in $\bm{\omega}/\omega_{1}\equiv\bm{\eta}$ yields
\begin{widetext}
\begin{equation}
\begin{aligned}
&h_{e}^{\bm{m}_{0}}=h^{\bm{m}_{0}}+\frac{1}{\omega_{1}}\sum_{n_{1}\neq0,\bm{n}_{0}}\frac{1}{2n_{1}}[h^{\bm{n}},h^{(-n_{1},\bm{m}_{0}-\bm{n}_{0})}]-\frac{1}{\omega_{1}}\sum_{n_{1}\neq0,\bm{n}_{0}}\frac{1}{n_{1}}[h^{\bm{n}},h^{\bm{m}_{0}-\bm{n}_{0}}]+\sum_{n_{1}\neq0}\frac{\bm{m}_{0}\cdot\bm{\eta}}{n_{1}}h^{(n_{1},\bm{m}_{0})}\\
&+\frac{1}{\omega_{1}}\sum_{n_{1}\neq0,\bm{n}_{0}}\frac{(\bm{m}_{0}+\bm{n}_{0})\cdot\bm{\eta}}{n_{1}^{2}}[h^{\bm{n}},h^{\bm{m}_{0}-\bm{n}_{0}}]+\frac{1}{\omega_{1}}\sum_{n_{1}\neq0,m_{1}\neq0,\bm{n}_{0}}\frac{(\bm{m}_{0}-\bm{n}_{0})\cdot\bm{\eta}}{m_{1}n_{1}}[h^{\bm{n}},h^{(m_{1},\bm{m}_{0}-\bm{n}_{0})}]\\
&+\frac{1}{2\omega_{1}}\sum_{n_{1}\neq\{0,m_{1}\},\atop m_{1}\neq0,\bm{n}_{0}}\frac{(m_{1}-n_{1})\bm{m}_{0}\cdot\bm{\eta}-m_{1}\bm{n}_{0}\cdot\bm{\eta}}{m_{1}n_{1}(m_{1}-n_{1})}[h^{\bm{n}},h^{(m_{1}-n_{1},\bm{m}_{0}-\bm{n}_{0})}].\label{MMFFE_first_order}
\end{aligned}
\end{equation}
Eq.~\eqref{MMFFE_first_order} then yields the explicit form of the coefficients
\begin{equation}
\begin{aligned}
C_{\Delta}&=\sum_{n_{1}\neq0}\frac{v(n_{1},0)-v(n_{1},-n_{1})/2}{n_{1}},\\
\tilde{C}_{\Delta}&=\frac{i}{\omega_{1}}\left(\sum_{n_{1}\neq0}\frac{3v(n_{1},0)}{n_{1}^{2}}+\sum_{n_{1}\neq0,\atop m_{1}\neq0}\frac{v(m_{1},n_{1})}{m_{1}n_{1}}+\sum_{n_{1}\neq\{0,m_{1}\},\atop m_{1}\neq0}\frac{v(n_{1},m_{1}-n_{1})}{m_{1}n_{1}}\right)\ .\\
D_{k}&=l_{k}^{(0)}-\frac{2\Delta}{\omega_{1}}\sum_{n_{1}\neq0}\frac{l_{k}^{(n_{1})}}{n_{1}},\\
\tilde{D}_{k}&=i\sum_{n_{1}\neq0}\frac{l_{k}^{(n_{1})}}{n_{1}}\left(1+\frac{4\Delta}{n_{1}\omega_{1}}\right),\\
E_{k}&=\sum_{n_{1}\neq0}\frac{p_{k}(n_{1},0)-p_{k}(n_{1},-n_{1})/2}{n_{1}},\\
\tilde{E}_{k}&=\frac{i}{\omega_{1}}\left(\sum_{n_{1}\neq0}\frac{3p_{k}(n_{1},0)}{n_{1}^{2}}+\sum_{n_{1}\neq0,\atop m_{1}\neq0}\frac{p_{k}(m_{1},n_{1})}{m_{1}n_{1}}+\sum_{n_{1}\neq\{0,m_{1}\},\atop m_{1}\neq0}\frac{p_{k}(n_{1},m_{1}-n_{1})}{m_{1}n_{1}}\right)\ ,
\label{Coe}
\end{aligned}
\end{equation}
\end{widetext}
with
\be
    v(n_{1},m_{1})=\sum_{k=1,2,3}l_{k}^{(-n_{1})\ast}l_{k}^{(m_{1})}-l_{k}^{(n_{1})}l_{k}^{(-m_{1})\ast}\ ,
    \ee
    and
    \be
    p_{k}(n_{1},m_{1})=l_{i}^{(m_{1})}l_{j}^{(-n_{1})\ast}-l_{i}^{(n_{1})}l_{j}^{(-m_{1})\ast}\ ,\label{v_and_p}
\ee
where the triple of indices $k,i,j$ adopts the values $1,2,3$ and the cyclic permutations $2,3,1$ and $3,1,2$.

The coefficients ${l_{k}^{(m)}}'s$ are the Fourier components of $\exp[i\bm{q}_{\text{lat}}(t)\cdot\bm{a}_{k}]$ with
\begin{equation}
    \bm{q}_{\text{lat}}(t)=\sum_{a=1}^{3}q_{a}\left[\sin(\omega_{a}t-\delta_{a})\bm{e}_{x}+\sin(\omega_{a}t-\delta_{a}^{\prime})\bm{e}_{y}\right].
\end{equation}
With the definitions in \eqref{Coe}, this yields
\begin{equation}
\tau_{1}=C_{\Delta}-\gamma\tilde{C}_{\Delta},\quad\tilde{\tau}_{k}=|E_{k}-\gamma\tilde{E}_{k}|,\quad\alpha_{k}=\mathrm{Arg}(E_{k}-\gamma\tilde{E}_{k}).\label{taus}
\end{equation}
\providecommand{\noopsort}[1]{}\providecommand{\singleletter}[1]{#1}%

\end{document}